\documentclass{article}
\usepackage{amsmath}
\usepackage{multirow}
\usepackage{PRIMEarxiv}
\usepackage{dblfloatfix}
\usepackage{float}
\usepackage[utf8]{inputenc} % allow utf-8 input
\usepackage[T1]{fontenc}    % use 8-bit T1 fonts
\usepackage{hyperref}       % hyperlinks
\usepackage{url}            % simple URL typesetting
\usepackage{booktabs}       % professional-quality tables
\usepackage{amsfonts}       % blackboard math symbols
\usepackage{nicefrac}       % compact symbols for 1/2, etc.
\usepackage{microtype}      % microtypography
\usepackage{lipsum}
\usepackage{fancyhdr}
\usepackage{svg}% header
\usepackage{array}
\usepackage{graphicx}
\usepackage{multirow}
\usepackage{enumitem}
\DeclareUnicodeCharacter{00A0}{ }
\newcommand\MyBox[2]{
  \fbox{\lower0.75cm
    \vbox to 1.7cm{\vfil
      \hbox to 1.7cm{\hfil\parbox{1.4cm}{#1\\#2}\hfil}
      \vfil}%
  }%
}
\usepackage{algorithm}
\usepackage{algpseudocode}
\usepackage{graphicx}       % graphics
\graphicspath{{media/}}     % organize your images and other figures under media/ folder

\usepackage{tikz,forest}
\usetikzlibrary{arrows.meta}
\forestset{
    default preamble={
        for tree={
            base=bottom,
            child anchor=north,
            align=center,
            s sep+=2.5cm,                      % or use: s sep +=1.5 cm ...
    straight edge/.style={
        edge path={\noexpand\path[\forestoption{edge},thick,-{Latex}] 
        (!u.parent anchor) -- (.child anchor);}
        },
    if n children={0}
        {tier=word, draw, thick, rectangle}
        {draw, diamond, aspect=2, thick}, % ... and retain: aspect=2
    if n=1{%
        edge path={\noexpand\path[\forestoption{edge},thick,-{Latex}] 
        (!u.parent anchor) -| (.child anchor) node[pos=.2, above] {Y};}
        }{
        edge path={\noexpand\path[\forestoption{edge},thick,-{Latex}] 
        (!u.parent anchor) -| (.child anchor) node[pos=.2, above] {N};}
        }
        }
     }
 }
\title{Neural Network and Order Flow, Technical Analysis: Predicting short-term direction of futures contract}

\author{
  Yiyang Zheng \\
  Department Of Engineering \\
  Shanghai University \\
  Shanghai, China \\
  \texttt{\ zhengyiyang@shu.edu.cn} \\
}

\begin{document}
\maketitle

\begin{abstract}
		Predictions of short-term directional movement of the futures contract can be challenging as its pricing is often based on multiple complex dynamic conditions. This work presents a method for predicting the short-term directional movement of an underlying futures contract. We engineered a set of features from technical analysis, order flow, and order-book data. Then, \textit{Tabnet}, a deep learning neural network, is trained using these features.  We train our model on the Silver Futures Contract listed on Shanghai Futures Exchange and achieve an accuracy of 0.601 on predicting the directional change during selected period. 
\end{abstract}

% keywords can be removed

\section{Introduction}
Like the equity market, the \textit{derivatives market }is often dynamic and volatile. Predicting the trend of a futures contract can be a challenging task since its pricing is often based on multiple complex dynamic conditions. Even though some critics claiming \cite{1-counterclaim} it is impossible to predict the financial market efficiently, researchers have demonstrated that complex algorithms may create predictive models that can out-perform random guessing \textit{consistently}.

Numerous methods have been proposed in this field. The more traditional method, including using time-series models like \textit{autoregressive integrated moving average }(ARIMA), has been proposed in papers \cite{1-arima} \cite{2-arima}. Those methods tend to over-fit on past data and yield a poor result on future unseen data. Nabipour et al. \cite{nabi} used various\textit{ technical indicators} to model the asset's return. In recent years, neural networks have been increasingly popular. Moghaddam et al. \cite{mogh} comparatively study the use of\textit{ Multilayer Perceptron }(MLP) and \textit{Recurrent Neural Networks} ( RNN )  on some unique features.

Some alternative methods include using news reports and social media are also suggested. Li et al. suggest using news sentiment to model the future movements of stocks \cite{li}, while \cite{1-sentiment-1}\cite{1-sentiment-2}\cite{1-sentiment-3}  suggesting using sentiment analysis and language processing to generate predictive features.

In this work, we present a method to predict the short-term directional movement of the futures contract. Our method involves a model that utilizes both \textit{technical analysis} features and \textit{order-flow} features that are generally applicable for both equities and derivatives and other features specifically designed for the futures contract.

The main contributions of our work are:

(1) We generate features from the\textit{ order book }and from\textit{ times and trades}. (2) We ensemble conventional technical analysis features and those generated from high-frequency book depth data ; (3)We apply\textit{ filters} to detect institutional traders from retail traders (4) The modeling of our features involved \textit{Purged Group Time Series Split} in contrast to traditional cross-validation to prevent target leakages and over-fitting (5)as a result, the accuracy of our model is not state-of-the-art, but the prediction accuracy is relatively \textit{consistent }across timespan regardless of market condition, which is overlooked in numerous of other works. (6) We can thus produce a high return on selected timespan with a basic trading strategy.

The remainder of this paper is structured as follows:

In the next section, we overview the problem and target construction. Then in section 3,4, we provide a method to generate features from the raw exchange data feed. Section 4 describes the design detail and section 5 focuses on training the model. In section 6, we provide a simple trading strategy and provide detailed performance statics of our model. Finally, in section 7, we conclude the research and identify a few possible directions for future research.

\section{Problem Statement}
We propose a general-applicable method to capture the trend of futures contracts through order-flow and technical analysis. To demonstrate and validate our method, we select Silver Futures Contract listed on Shanghai Futures Exchange, one of the most liquidated precious metal futures contracts in the world, with more than 30 billion yuan traded every session \textbf{Table 1 }compare this contract annual volume with similar contracts listed on other exchange.
\newline
\begin{table}[htb]
  \centering
\caption{Comparison in Annual Volume of Silver Contract Across Exchange}
\begin{center}

  \begin{tabular}{  c | c }
    \toprule

   \textbf{  Exchange}&\textbf{Annual Volume of 2021(In thousands of Contracts)}\\
      \hline

   Shanghai Futures Exchange Silver &\textbf{ 231,457}\cite{sfevolume}\\
   COMEX Silver &  26,126\cite{cmevolume}\\    
    \bottomrule

  \end{tabular}
\end{center}
\label{volumecompare}
\end{table}

We selected the near-month silver contract listed on Shanghai Futures Exchange from January 1 st,2018, to Dec 31st,2021. This time frame reflects various market conditions, including a lengthy, less-volatile timeframe and increased volatility during the coronavirus pandemic, thus reflecting a general market condition and preventing over-fitting. \textbf{Fig 1 }shows the price of the front-month futures contract during the period. Shanghai Futures exchange does not offer an order-based market feed but instead provides a 500-millisecond snapshot of trades through a propriety protocol, providing order-book, aggregated trade, and open interest of the contract. We utilized an unfiltered feed to create features. 

\begin{figure}
  \centering
  \includegraphics[scale=2]{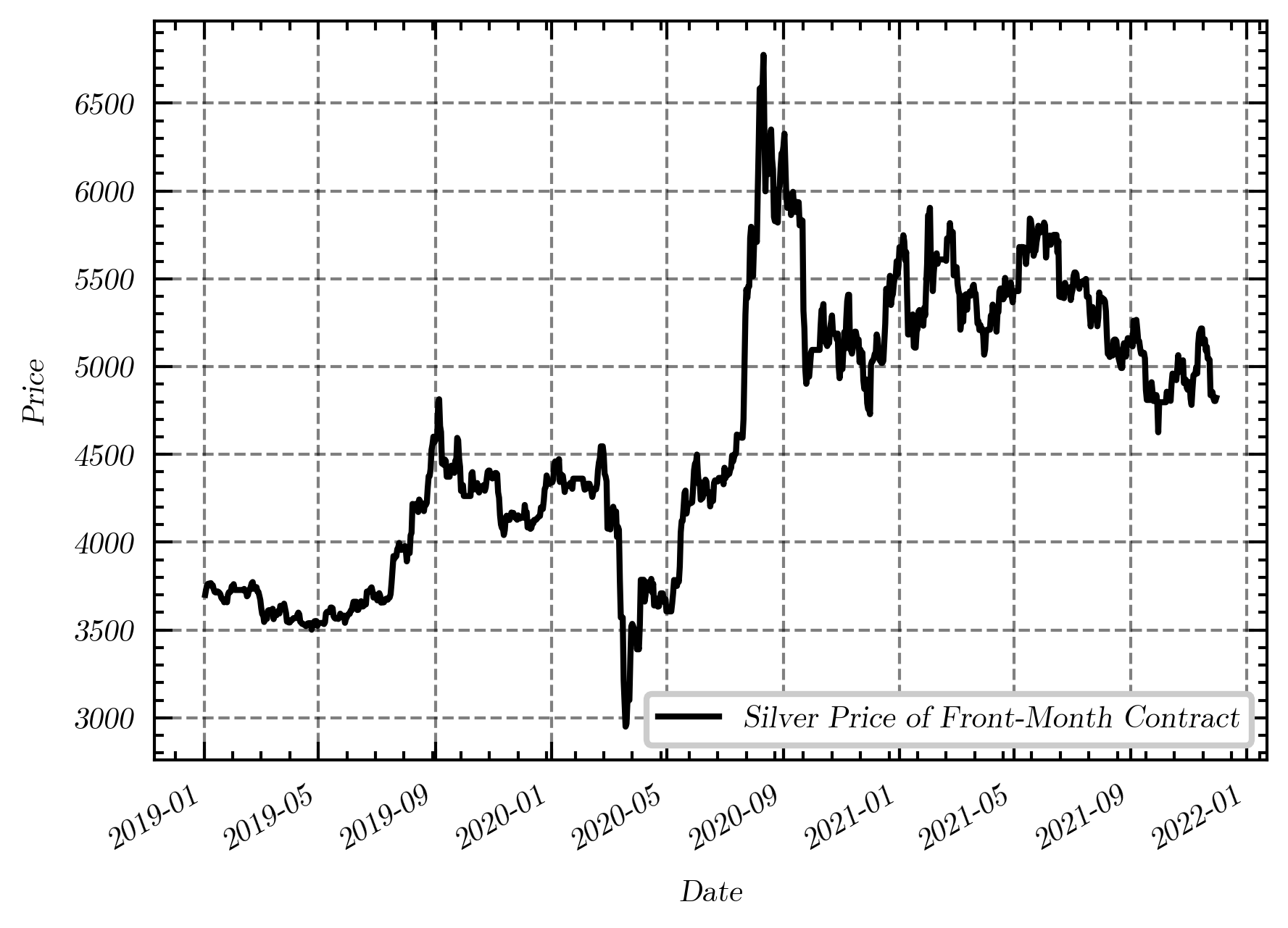}
  \caption{\textit{Silver Price of Front-Month Contract during January 1 st,2018 to Dec 31 st,2021}}
\end{figure}
Our goal is to predict the direction of movement on this particular contract in the next 15 minutes. To make our target more stable, we apply 2 minutes moving window and calculate VWAP to smooth the ultra-short-term spike, as defined below:
\[P^\alpha\left(t\right)=\frac{\langle Psession*Vsession \rangle}{\langle Psession \rangle}\]
\begin{center}
\textit{where \(P^\alpha\left(t\right)\) represents rolling average of 120 snapshots (~2 minutes)}
\end{center}
Then, we calculate the log return based on this smooth price:

\[R^\alpha(t)\ =\ \ log(\frac{P^\alpha(t+15)}{P^\alpha(t)})\]
Finally, we calculate \({\rm Target}^\alpha(t)\) based on \(R^\alpha(t)\) .

\[{\rm Target}^\alpha(t)=
\begin{cases}
1, & P^\alpha\left(t\right)>0,\\
0, & P^\alpha\left(t\right)< 0
\end{cases}\]
Since it is unlikely that volume-weighted price remained unchanged during the 15 minutes timespan, we ignore target where \( R^\alpha(t)\ =\ 0\).

Another consideration is the data-point with minimum movement .\(R^\alpha(t)\) with only a few tick changes can create noisy training targets and are not practical to utilize in actual market conditions. To resolve this, we  remove those data-point\( {|R}^\alpha(t)| < \theta\\\)  In our case \( \theta\ =\ 0.001\). The total samples decreased from 38 ,294 ,652 before processing to 24 ,347, 353.

As shown in \textbf{Table 2}, the labels are evenly distributed across two labels during the time frame.
\begin{table}[H]
  \centering
\caption{\textit{Target Distribution}}
\begin{center}

  \begin{tabular}{ c | c |c}
    \toprule

  \textbf{   Target}&\textbf{Sample}&\textbf{Percentage(\%)}\\
     
      \hline

   1 &12 ,389 ,465&50.88\\
  0 &  11 ,957 ,888&49.11\\    
    \bottomrule

  \end{tabular}
\end{center}
\end{table}
\section{Data and pre-processing}

\subsection{Market Data Structure}
Shanghai Futures Exchange provides a snapshot-based order feed using\textit{ CTP} protocol. The feed aggregates the change during the last 500 milliseconds. Each feed contain multiple fields, including the\textit{ trade} and \textit{order-book} information; we utilize the following fields:
\begin{enumerate}
    \item\(  price_{i}\)\newline
           Last traded price at snapshot \textit{i}
    \item\(  volume_{i}\)\newline
           Accumulated volume of the futures contract at snapshot \textit{ i }
    \item\(  openInterest_{i}\)\newline
           Open Interest of the futures contract at snapshot \textit{ i }
    \item\( {\rm sizebid}_i^n\)\newline
            Size of \textit{n} -th level of order-book at snapshot  \textit{ i } on bid-side
    \item\( {\rm sizeask}_i^n\)\newline
            Size of \textit{n} -th level of order-book at snapshot \textit{ i }  on ask-side
      \item\( {\rm pricebid}_i^n\)\newline
       		 Quoted price of \(n\)-th level of order-book at snapshot \textit{i } on bid-side    \item\( {\rm priceask}_i^n\)\newline
       		 Quoted price of \(n\)-th level of order-book at snapshot \textit{i } on ask-side
\end{enumerate}
\subsection{Market Data Pre-processing}

From the original market feed, we calculate several features for the convenience of  further feature engineering. Those features are not directly used in our final model.
\subsubsection{Volume Change and Open Interest Change}
\begin{itemize}
\item \textbf{Open Interest and Volume change}

We calculate the Open Position Change for the snapshot \textit{ i} using.

\[{\rm openInterestChg}_i\ =\ {\rm openInterest}_{i\ -\ }{\rm openInterest}_{i\ -\ 1}\]

\item \textbf{Volume Change}

We calculate the Volume Change for the snapshot \textit{ i} using.

\[{\rm volumeChg}_i\ =\ {\rm volume}_{i\ -\ }{\rm volume}_{i\ -\ 1}\]
\end{itemize}
\subsubsection{Open Contracts and Close Contracts}
Since we already calculate the change of open interest and volume and for all timestamp,for a given contract,we can calculate the contracts opened and closed at snapshot  \textit{ i }using:
\begin{equation}
\begin{cases}
&  openContract_{i} + closeContract_{i} = volumeChg_{i}\\
&  openContract_{i} - closeContract_{i} = openIntrestChg_{i}
\end{cases}
\end{equation}
\begin{itemize}
\item \textbf{Open Contracts}

\end{itemize}
\[{\rm openContract}_i\ =\ \ \frac{{\rm openInterest}_i\ +\ {\rm volume}_i}{2}\]
\begin{itemize}
\item \textbf{Close Contracts}

\[{\rm closeContract}_i\ =\ {\rm openInterest}_i\ -\ {\rm volume}_i\]
\end{itemize}
\subsubsection{Snapshot Type }
We catalog snapshot  \textit{ i}  into several types based on directional change of price and open interest, using \textbf{Figure 2}:

\begin{figure}[H]
 \centering

    \includegraphics[width=0.7\columnwidth]{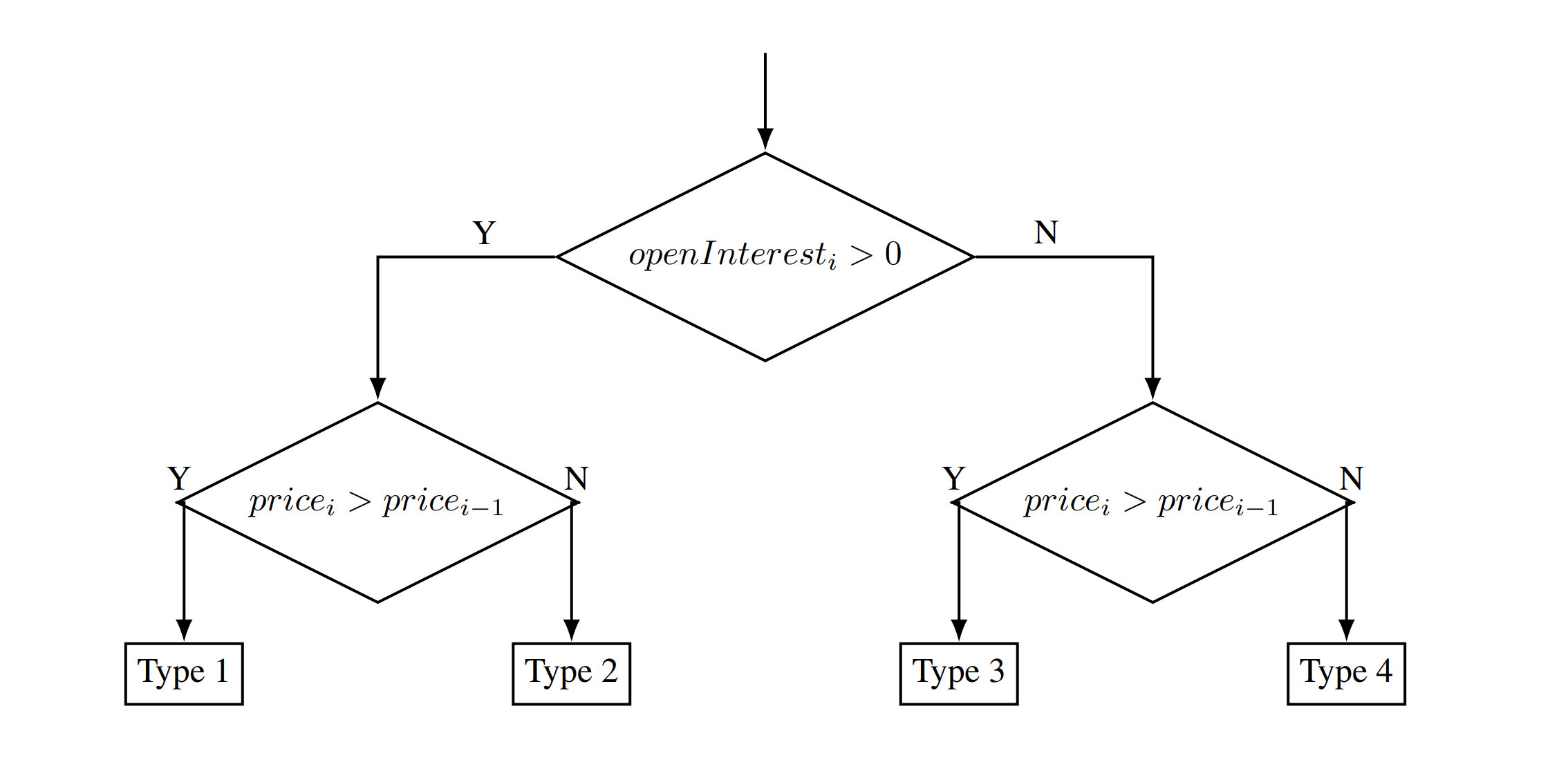}
    \caption{Snapshot Classification}
  \label{fig:fig1}
\end{figure}

\begin{itemize}
\item \textbf{Type 1}
\newline
Type 1 snapshot indicates that the snapshot price is \textit{higher }than the previous snapshot and open interest \textit{increase}, presumably caused by market participants taking a\textit{ long} position\textit{ increasing} their position.
\item \textbf{Type 2}
\newline
Type 2 snapshot indicates that the snapshot price is \textit{lower} than the previous snapshot and open interest\textit{ increase}, presumably caused by market participants taking a\textit{ short} position \textit{increasing} their position.
\item \textbf{Type 3}
\newline
Type 3 snapshot indicates that the snapshot price is \textit{higher} than the previous snapshot and open interest \textit{decrease}, presumably caused by market participants taking a \textit{long }position\textit{ decreasing} their position.
\item \textbf{Type 4}
\newline
Type 4 snapshot indicates that the snapshot price is\textit{ lower} than the previous one, and open interest \textit{decreases}, presumably caused by market participants taking a \textit{short }position and \textit{decreasing} their position.
\end{itemize}

\subsubsection{OHLCV(Open-High-Low-Close-Volume) Bar}

An \textit{OHLCV} chart is a bar chart that shows open, high, low, and closing prices and overall volume for each period. We aggregate the bar every 1 minute (120 snapshots) in our usage. \textit{OHLCV }charts are helpful since they show the data change over a period, and we can generate technical analysis features utilized by the model.

\section{Feature Engineering}
This section provides details of the generation of features utilized in our model.

\subsection{Technical Analysis Feature}
\textit{Technical analysis}, unlike fundamental analysis, attempts to find the pattern of price and volume, modeling supply and demand to predict price movement. Moreover, technical analysis can generate short-term trading signals. There is an extensive range of technical analyses; we utilize some of them using  a 1-minute OHLCV bar generated from snapshot as described in section\textbf{ 3.2.4}. We use all the indicators with parameters specified in the \textbf{Table 3, Table 4, Table 5, Table 6 } as features.

\begin{table}[H]
\centering
\caption{\textit{Technical Analysis} Volume Indicators}
  \begin{tabular}{c|c}

    \toprule

\textbf{     Indicator }    &\textbf{ Parameter}  \\
    \midrule
    Acc. Dist. Index Indicator & \multirow{3}{*}{\textit{None}}    \\
    On Balance Volume Indicator &  \\
    Chaikin Money Flow Indicator        &\\
      \hline
    Force Index Indicator &\( window=13\) \\
    Ease Of Movement Indicator &\(window=14\)\\
    Volume Price Trend Indicator &\(window=14\)\\
    
    \bottomrule

  \end{tabular}

\end{table}

\begin{table}[H]
  \centering
\caption{\textit{Technical Analysis} Volatility Indicators}
  \centering
  \begin{tabular}{c|c|c}
    \toprule

 \textbf{   Base}     & \textbf{Indicator}     &\textbf{ Parameter } \\
    \midrule
     \multirow{5}{*}{Bolling Bands}& $P-Band$&   \multirow{5}{*}{\(window=20, windows_{dev}=2\)}  \\
        & $ W-Band$  & \\
  	
     &$Higher-Band$ &  \\
      &$Middle-Band $&  \\
    &$ Lower-Band$&\\

   \hline
       \multirow{5}{*}{Keltner Channel}&$ W-Band$&   \multirow{5}{*}{\(window=10\)}  \\
     &$C-Band$&  \\
     &$P-Band$&\\
    & $Higher-Band$ &\\
   &$ Lower-Band$& \\

    \hline
       \multirow{5}{*}{Donchian Channel}&$C-Band$ &   \multirow{5}{*}{\(window=20\)}  \\
     &$P-Band$&  \\
        & $W-Band$ & \\
        & $Lower-Band$&\\
    & $Middle-Band$&\\

    \bottomrule

  \end{tabular}

\end{table}
\begin{table}[H]
  \centering
\caption{\textit{Technical Analysis} Trend Indicators}
  \centering
  \begin{tabular}{c|c|c}
    \toprule

 \textbf{   Base}     & \textbf{Indicator}     &\textbf{ Parameter } \\
    \midrule
     \multirow{3}{*}{MACD}& $MACD$&   \multirow{3}{*}{\({ window_{slow}=26, window_{fast}=12, window_{sign}=9}\)}  \\
     &$MACD_{Signal}$ &  \\
    &$ MACD_{Diff}$ &\\
    \hline
     \multirow{2}{*}{SMA}& $SMA_{Fast}$&\( window=16\)\\
     &$SMA_{Slow}$ &\( window=32\)\\

    \bottomrule

  \end{tabular}

\end{table}
\begin{table}[H]
  \centering
\caption{\textit{Technical Analysis} Momentum Indicators}
  \centering
  \begin{tabular}{c|c|c}
    \toprule

\textbf{    Base  }   & \textbf{Indicator  }   &\textbf{ Parameter}  \\
    \midrule
     \multirow{3}{*}{ $Stoch RSI$}&  $StochRSI$&   \multirow{3}{*}{ \(window_{slow}=26, window_{fast}=12, window_{sign}=9\)}  \\
     & $StochRSI_K$ &  \\
    &  $StochRSI_D $ &\\

    \bottomrule

  \end{tabular}

\end{table}

\subsection{Order Book Features}
\textit{Order book }refers to an electronic list of buy and sell orders for a specific security or financial instrument. The order book provides detailed information about the sale–buy strength. 
\begin{itemize}
\item Bid-Ask Spread
\newline
Accumulated spread at order-book level k at snapshot i  ,using :
\[{\rm AccumulatedSpread}_i^k\ =\ \sum_{n\ =\ 1}^{k}{{\rm priceask}_i^n\ -\ }\ {\rm pricebi d}_i^n\]
Then, a\textit{ rolling window} is applied to represent order-book overall statics over the last \textit{m} minutes(\textit{120*m} snapshots).
\[{\rm AverageAccumulatedSpread}_i^k\ =\ \frac{1}{120m}\sum_{i-120m}^{i}{\rm AccumulatedSpread}_i^k\ \]
For each snapshot , we calculate accumulated spread of top 5 levels (k=1,2,3,4,5) ,and rolling period\textit{ m} of 5,10,15,30 minutes thus generating\textit{ 20 }features 
\item  Volume imbalance
\newline
Accumulated spread at order-book level k at snapshot i ,using :
\[{\rm AccumulatedVolumeImbalance}_i^k\ =\ \sum_{n\ =\ 1}^{k}{{\rm sizeask}_i^n\ -\ }\ {\rm sizebi d}_i^n\]

			Then, a rolling window is applied to represent order-book overall statics over the last \textit{m} minutes(\textit{120m} snapshots).
\[{\rm AverageAccumulatedVolumeImbalance}_i^k\ =\ \frac{1}{120m}\sum_{i-120m}^{i}{\rm  VolumeImbalance}_i^k\ \]
For each snapshot , we calculate volume imbalance of top 5 levels (k=1,2,3,4,5) ,and rolling period \textit{m} of 5,10,15,30 minutes thus generating \textit{20} features .

\end{itemize}
\subsection{Order Type Features}
As specified in section\textbf{ 3.2.3} ,we classify the snapshot into 4 catalogues.We generate features according to the percentage of each type during the last \textit{m} minutes(\textit{120*m} snapshots).
\[{\rm SnapshotTypePct}_i^k\ =\ \frac{1}{\left|S\right|}\sum_{snapshot\in{S}}{1_{{SnapshotType\ =\ k}}(snapshot)}\]
\begin{center}

W\textit{where S represents a set of snapshots from i -120m to i .}
\end{center}
Further, we want to filter snapshots where larger orders take place,presumably placed by institution investors,thus,we have : 
\[{\rm SnapshotTypePctFilter}_i^k\ =\ \frac{1}{\left|S\right|}\sum_{snapshot\in{S}}{1_{{SnapshotType\ =\ k}}(snapshot)}\]
\begin{center}

\textit{where S represents a set of snapshots from i -120m to i, excluding snapshots with volume change more minor than \textbf{10 }contracts.}
\end{center}

We calculate all four snapshot types with a rolling period\textit{ m} of 5,10,15,30 minutes for each snapshot, thus generating \textit{32} features.
\subsection{Order Flow features}
In session \textbf{3.2.1 }and \textbf{3.2.2}, we derived data such as opened contracts and closed contracts from original snapshots. Utilizing those data, we calculate the following features :
\begin{itemize}
\item Open Close Percentage 
\newline
Percentage of closed contracts in ratio to opened contracts.
\[ {\rm openClosePct}_i = \frac{\sum_{k\ =\ i\ -\ 120m}^{i}{\rm openContract}_k}{\sum_{k\ =\ i\ -\ 120m}^{i}{\rm closeContract}_k} \]
\item Open Interest Change\newline
Percentage of open interest change.
\[ {\rm openInterestChg}_i\ =\ \frac{{\rm openInterest}_i}{{\rm openInterest}_{i-120m}}\]
\end{itemize}

We calculate a rolling period\textit{ m} of 5,10,15,30 minutes for each snapshot, thus generating 8 features.
\subsection{Correlation Analysis}
We test the correlations between the features and the target before building the model,we perform a Pearson Correlation \cite{pearson} between each feature and target,as defined below:
\begin{equation*}
  r =
  \frac{ \sum_{i=1}^{n}(x_i-\bar{x})(y_i-\bar{y}) }{%
        \sqrt{\sum_{i=1}^{n}(x_i-\bar{x})^2}\sqrt{\sum_{i=1}^{n}(y_i-\bar{y})^2}}
\end{equation*}
\begin{center}

\textit{where x is the feature and y is the target}
\end{center}

The correlation distribution of all the features is shown in\textbf{ Table 7}

\begin{table}[H]
\centering
\caption{\textit{Feature Correlation Distribution}}
  \begin{tabular}{c|c}

    \toprule

\textbf{Metrics}    &\textbf{ Value}  \\
    \midrule

  	mean & 0.017307\\
    std &	0.080738\\
   max& 0.204487\\
    25\% & 0.067418\\
    50\% & 	0.015707\\
    75\% & 	0.019624\\
    min & 	-0.02934\\
    \bottomrule

  \end{tabular}

\end{table}

\section{Model}
\subsection{Model Selection}
We utilize \textbf{TabNet} as our model. \textbf{TabNet} uses sequential attention to choose which features to reason from at each decision step, enabling interpretability and more efficient learning as the learning capacity is used for the most salient features. \cite{tabnet}
\subsection{Metric}
We utilize \textbf{AUC-ROC} (Area Under the Receiver Operating Characteristics ) as our metric.AUC - ROC curve is a performance measurement for the classification problems at various threshold settings. ROC represents a probability curve, and AUC represents separability. Our train data may contain a period of unbalanced labels, so using only accuracy or recall score is inappropriate.\newline
To calculate AUC-ROC, we first calculate \textit{ true/false positive} and\textit{ true/false negative} instances.
A\textit{ true positive} occurs when the model correctly predicts the positive class. Similarly, a \textit{true negative} is an outcome where the model correctly predicts the negative class. A \textit{false positive} occurs when the model incorrectly predicts the positive class. Moreover, a \textit{false negative }is when the model incorrectly predicts the negative class.
Then,we calculate specificity and sensitivity using the formula:

\[Specificity\ =\ \frac{True\ Negatives\ }{True\ Negatives\ +\ False\ Positives}\]
\[Sensitivity\ =\ \frac{\ True\ Positives}{True\ Positives\ +\ False\ Negatives}\]
Finally, we draw the curve using different thresholds and calculate the area under \textit{ Sensitivity }and \textit{1 - Specificity}.
\subsection{Hyper-parameter}
Hyper-parameters are a set of parameters that control the learning process and model structure. We do not specifically search for optimal hyper-parameter as it may lead to over-fitting, but we specify some parameters to control the width and depth of the model. 
\begin{table}[H]
\centering
\caption{\textit{Model Hyper-parameter}}
  \begin{tabular}{c|c|c}

    \toprule

\textbf{ Parameters}    &\textbf{ Value}    &\textbf{ Description}  \\
    \midrule

    \( n_d\) & 32&Width of the decision prediction layer. \\
  \(  n_a\)&32&	Width of the attention embedding for each mask.\\
\(   n_{steps}\)&5&Number of steps in the architecture\\

    \bottomrule

  \end{tabular}

\end{table}
\subsection{Cross-Validation}
\textit{Cross-validation} is a method that splits the dataset into different parts and then uses a different fraction for training and validation. Cross-validation can effectively protect overfitting, specifically over a small or diverse dataset. Despite having multi-million rows of the dataset, the future data tend to be volatile in our instance. Thus, a single model will easily overfit the training dataset.\newline
However, the conventional cross-validation method carries the risk of overfitting. Since each of our training instances is not isolated but a time-series, without a gap, some features having some lag or window calculations in them may leak unseen information. Thus, we must introduce a gap to prevent the risk of leaking information.\newline
We utilize \textit{Purged Group Time Series Split } for our model. In  \textit{Purged Group Time Series Split}, test indices must be higher than before in each split without any shuffling. A gap is introduced between the training and validation datasets to deal with lagged features. We split the dataset into five folds. A demonstration of the split is provided below:

\begin{figure}[H]
        \centering
        \includegraphics[width=0.7\columnwidth]{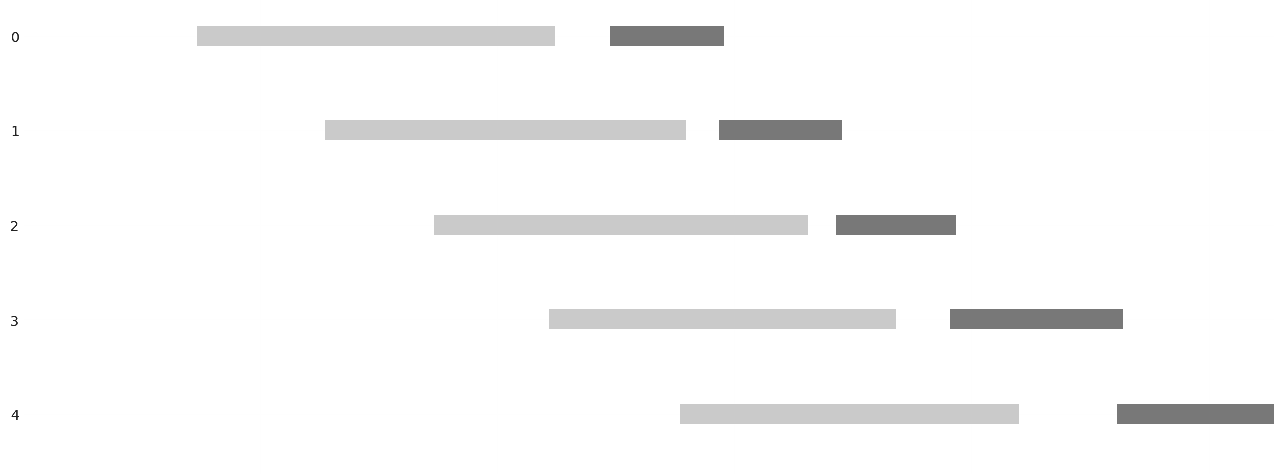}
          \caption{Demonstration of  \textit{Purged Group Time Series Split} }
    \end{figure}
\subsection{Training}
We train the model on the dataset with dates ranging from Jan 1st,2018 to Aug 31st,2021, with a sample size of 21,642,091. We perform a semi-supervised pre-training before training. The training curve of the model is shown below.
\begin{figure}[H]
\centering
\begin{minipage}{.6\textwidth}
  \centering
  \includegraphics[width=.5\linewidth]{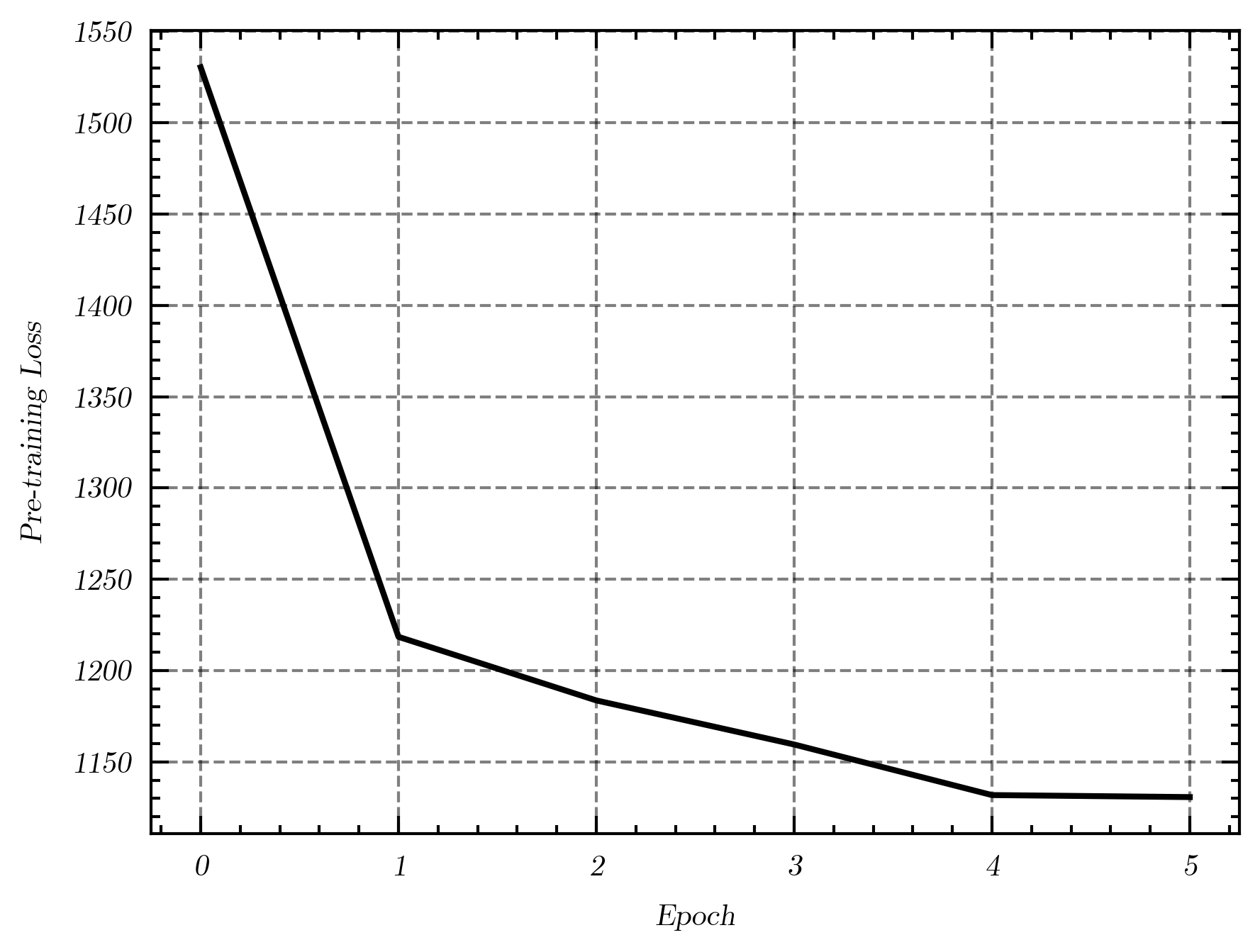}

\end{minipage}%
\begin{minipage}{.6\textwidth}
  \centering
  \includegraphics[width=.5\linewidth]{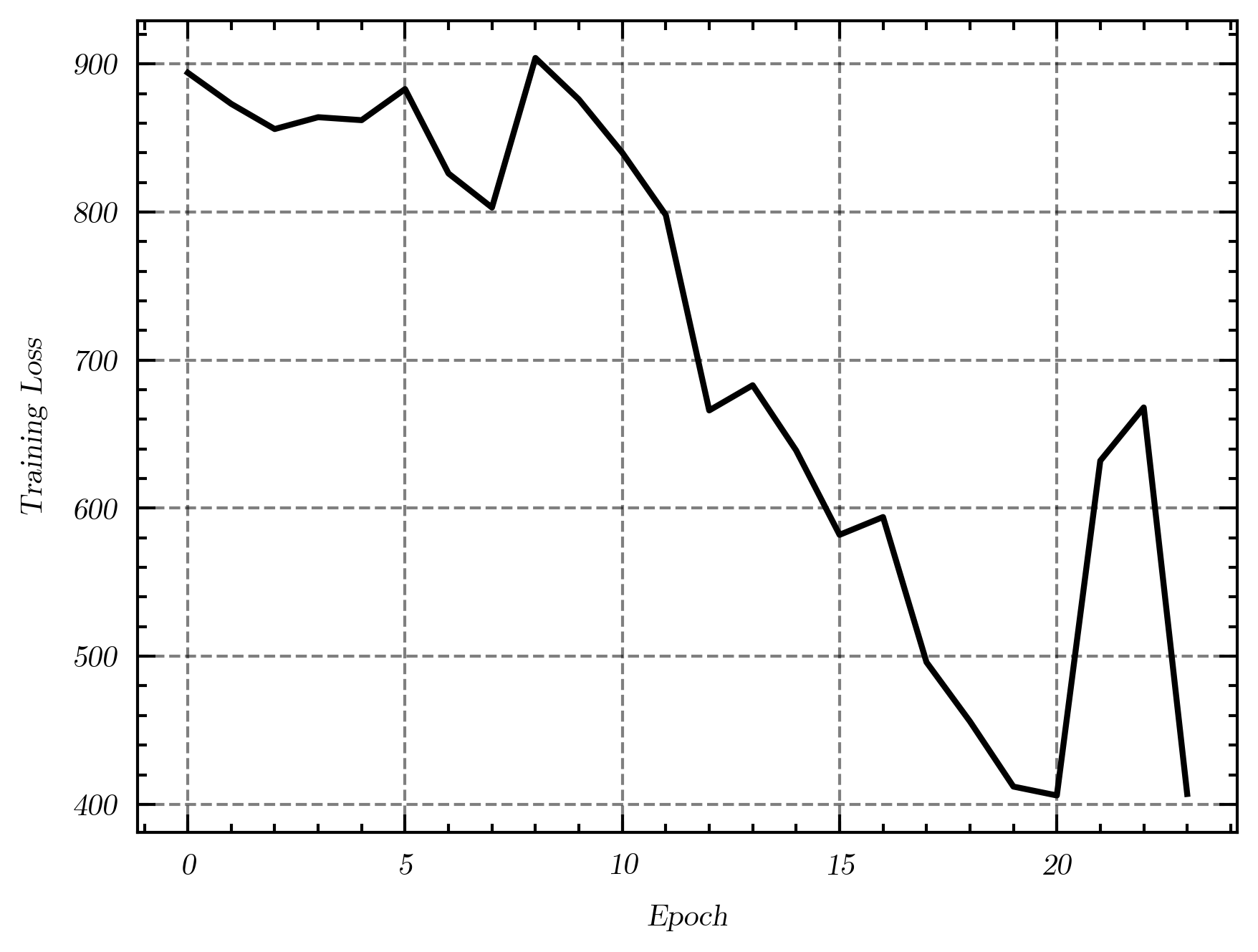}

\end{minipage}
  \centering
\caption{Pre-Training and Training Curve }
\end{figure}
\section{Model Validation and Backtesting}
In this section, we first validate the model on a  test dataset, then with a basic trading strategy, we examine the model's profitability.
\subsection{Validation Method}
We validate the  model on the dataset with dates ranging from Sept 1st,2021, to Dec 1st,2021. The label distribution of the test dataset is shown in \textit{Table 9}.
\begin{table}[H]
  \centering
\caption{Target Distribution}
\begin{center}

  \begin{tabular}{  c | c |c}
    \toprule

  \textbf{   Target}&\textbf{Samples}&\textbf{Percentage(\%)}\\
     
      \hline

   1 &695 ,200&49.35\\
  0 & 713 ,298&50.64\\    
    \bottomrule

  \end{tabular}
\end{center}
\end{table}
Since we perform a \textit{Cross-validation} on the dataset , we have a total of 5 models , we first individually predict the \textit{probability } of target to be 1 for each model , then ensemble the result using:
\[pred = \frac{\sum_{i\ \ =\ 1}^{5}{{model\_pred}_i\ }-\ max(model\_pred)\ -\ min(model\_pred)}{3}\]
We then convert the label predicted using a threshold of 0.5.
\subsection{Validation Result}
We compare the label predicted with the ground truth. The confusion matrix is shown below :
\begin{figure}[H]
        \centering
        \includegraphics[width=0.7\columnwidth]{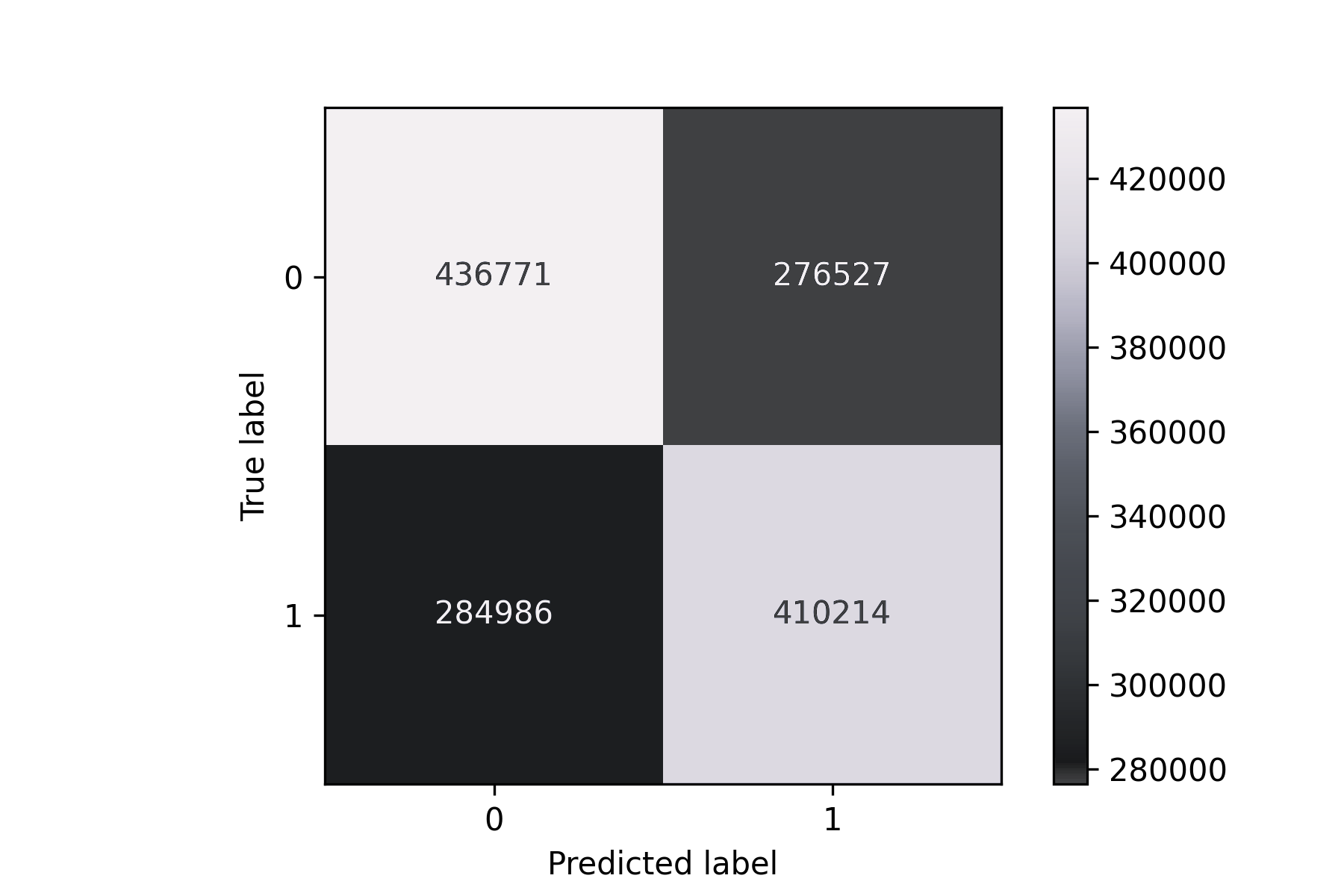}
          \caption{ \textit{Confusion Matrix} }
    \end{figure}

We further calculate some metrics on the prediction result.
\begin{table}[H]
\centering
\caption{\textit{Performance Metrics}}
  \begin{tabular}{c|c}

    \toprule

\textbf{Metrics}    &\textbf{ Value}  \\
    \midrule

  	Accuracy & 0.601339\\
    Recall & 0.590069\\
   Pearson Correlation& 0.207273\\

    \bottomrule

  \end{tabular}

\end{table}
\subsection{Backtesting}

	We perform backtesting to test the profitability of our model. To simplify our simulation, we make the following assumptions:
\begin{itemize}
\item We always hold either 100\% cash or 100\% futures contract.
\item We ignore the fee charged by the exchange and broker.
\item We assume all our orders are filled precisely at the last traded price.
\item We utilize the minimum margin ratio (10\%) required by the exchange.
\end{itemize}
	
\begin{algorithm}
	\caption{\textit{Trading strategy}}
	\label{tradingalgo}
	\begin{algorithmic}[]
		
		\State \textbf{Input:} Model predictions  probabilities $\hat{y}_1,\ldots,\hat{y}_T$, account holding state $PX$, threshold $\gamma \geq 0$
        \State $ {PX \gets None} $
        \State${ \gamma \gets 0.25}$
		\For{$t=1,\ldots,T$}		\Comment{Loop over each 15 minute interval}
		\If{$PX = None$}	\Comment{Currently holding no position}
		\If{$\hat{y}_t \geq (0.5+\gamma$)}
          \State\textbf{Open Long Position}  \Comment{We only open position if the prediction  is confident enough.}
          \State $ {PX \gets Long}$
         \EndIf
        \If {$\hat{y}_t \leq (0.5-\gamma$)}
          \State\textbf{Open Short Position}
          \State $ {PX \gets Short}$

		\EndIf
		\Else 	\Comment{Currently holding position}
		\If{PX = Long AND $\hat{y}_t \leq 0.5$} \Comment{The prediction has reversed,we close position.}
        \State\textbf{{Close Long Position}}
             \State $ {PX \gets None}$
        \EndIf
         \If{PX = Short AND $\hat{y}_t \geq 0.5$}
        \State\textbf{{Close Short Position}}
         \State $ {PX \gets None}$
        \EndIf
        \EndIf
		\EndFor
  
	\end{algorithmic}
\end{algorithm}
We simulate the trading using the same dataset used for validation and the  \textbf{Algorithm 1}. The total return and price change during the period is shown in\textbf{ figure 6}. 
\begin{figure}[H]
        \centering
        \includegraphics[width=0.7\columnwidth]{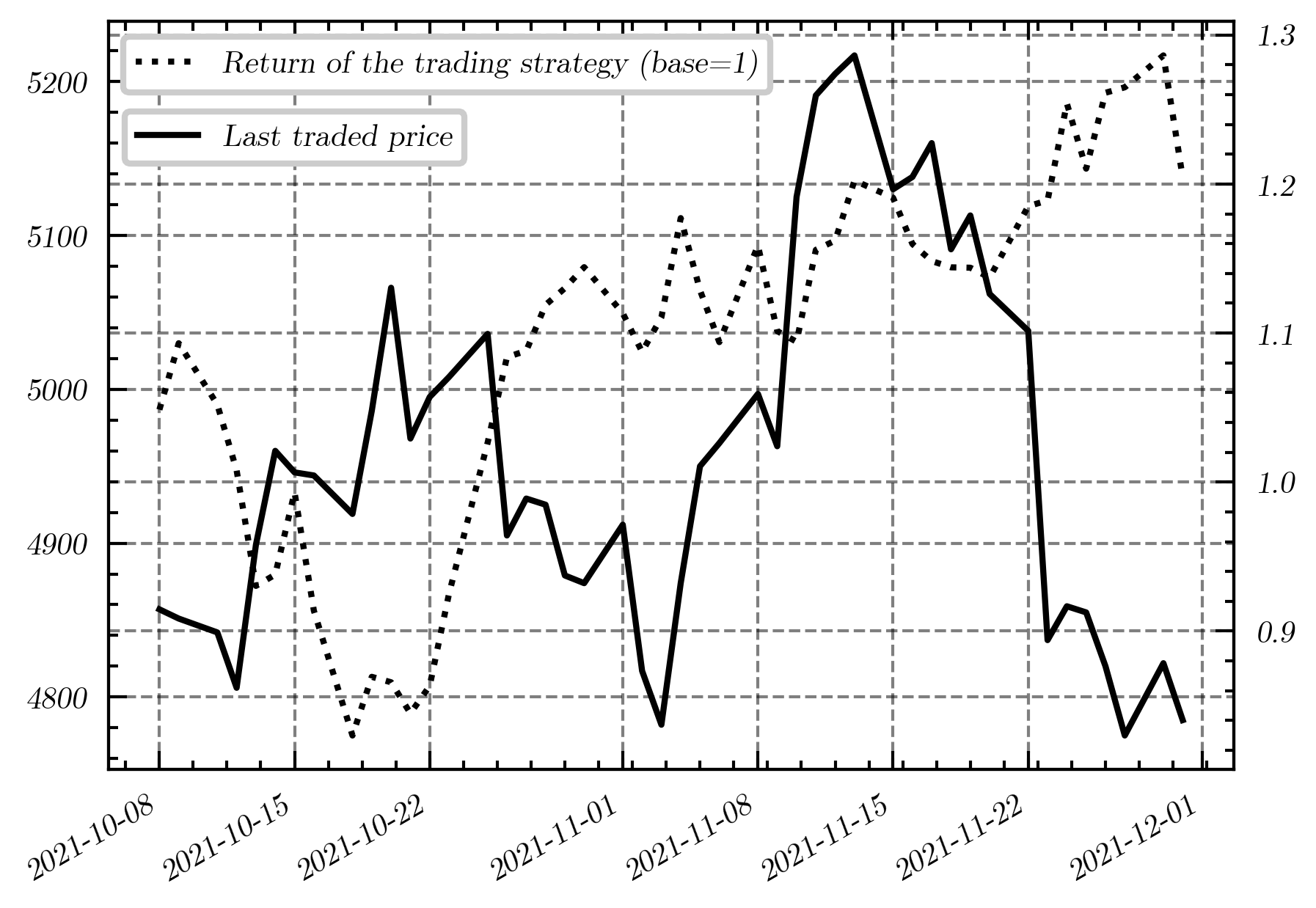}
          \caption{ \textit{Strategy Return and Price of the Contract} }
    \end{figure}
Despite the  futures contract price remaining nearly unchanged during the period, our model can still generate a relatively decent return during the period. We further calculate some common metrics to measure the model's performance.
\begin{table}[H]
\centering
\caption{\textit{Performance of the strategy}}
  \begin{tabular}{c|c}

    \toprule

\textbf{ Metrics}    &\textbf{ Value}  \\
    \midrule

  	Max drawdown & 24.08\%\\
    Annualized Sharpe Ratio& 1.2054\\
   Total Return& 20.48\%\\

    \bottomrule

  \end{tabular}

\end{table}
The model can achieve a decent return, but the daily performance tends to be volatile, which may be attributed to the  high margin ratio we used in the simulation.

\section{Conclusion and Discussion}
This paper discussed a neural network model using multiple features generated from conventional technical analysis and market micro-structure like order flow and order book. We train these models using a specified cross-validation method. When ensembled, the model is profitable even with a basic trading strategy. Thus, we conclude that this method is sufficient to gain insight into the short-term directional change of the futures market.
\subsection{Possible Extensions}
\begin{itemize}
\item Feature Engineering \newline
We may utilize alternative data, such as news and social media data, to generate features like sentiment scores.
\item Model \newline
We may ensemble a non-neural-network model, for instance, a tree-based model with a neural network model to gain better accuracy.
\end{itemize}

%Bibliography
\bibliographystyle{unsrt}  
\bibliography{references}

\end{document}